\documentclass[english,fleqn,twoside]{article}
\usepackage[T1]{fontenc}
\usepackage[latin1]{inputenc}
\usepackage{amsmath}
\usepackage{amssymb}

\makeatletter
\usepackage[headings]{espcrc2}

\title{\vspace{-2cm}Derived Brackets from Super-Poisson Brackets}

\author{S. Guttenberg\address{Institut f\"ur Theoretische Physik, TU Wien}\thanks{The author would like to thank the organizers for the great Carg\`ese summer school and for funding. The present support by the FWF project P19051 is also acknowledged.}}
       
\runtitle{Brackets}
\runauthor{S. Guttenberg}

\usepackage{babel}
\makeatother
\begin{document}
\newcommand{\bs}[1]{\boldsymbol{#1}}
\newcommand{\mf}[1]{\mathfrak{#1} }
\newcommand{\mc}[1]{\mathcal{#1}}
\newcommand{\dann}{\Rightarrow}
\newcommand{\hoch}[1]{{}^{#1 }}
\newcommand{\tief}[1]{{}_{#1 }}
\newcommand{\erw}[1]{\langle#1\rangle}
\newcommand{\Erw}[1]{\left\langle #1 \right\rangle }
\newcommand{\basis}{\boldsymbol{\mf{t}}}
\newcommand{\ip}{\imath}
\newcommand{\de}{{\bf d}\!}
\newcommand{\es}{{\bf s}\!}
\newcommand{\dew}{\bs{d}^{\textrm{w}}\!}
\newcommand{\Lie}{\bs{\mc{L}}}
\newcommand{\Dorf}{\bs{\mc{D}}}
\newcommand{\pe}{\bs{\partial}}
\newcommand{\one}{1\!\!1}
\newcommand{\lqn}[1]{\lefteqn{#1}}
\newcommand{\partiell}[2]{\frac{\partial#1 }{\partial#2 }}
\newcommand{\Partiell}[2]{\left( \frac{\partial#1 }{\partial#2 }\right) }
\newcommand{\ola}[1]{\overleftarrow{#1}}
\newcommand{\lpartial}{\overleftarrow{\partial}}
\newcommand{\partl}[1]{\frac{\partial}{\partial#1}}
\newcommand{\partr}[1]{\frac{\lpartial}{\partial#1}}
\newcommand{\funktional}[2]{\frac{\delta#1 }{\delta#2 }}
\newcommand{\funktl}[1]{\frac{\delta}{\delta#1}}
\newcommand{\funktr}[1]{\frac{\ola{\delta}}{\delta#1}}
\newcommand{\abs}[1]{\mid#1 \mid}
\newcommand{\be}{\bs{b}}
\newcommand{\ce}{\bs{c}}
\newcommand{\tet}{\bs{\theta}}
\newcommand{\Es}{\bs{S}}
\newcommand{\To}{\rightarrow}

\begin{abstract}
The relation between Poisson brackets in supersymmetric one or two-dimensional
sigma-models and derived brackets is summarized. \vspace{1pc}
 
\end{abstract}
\maketitle 
\begin{picture}(0,0)\unitlength=1mm\put(145,70){TUW--07--03}\end{picture}

Alekseev and Strobl observed in \cite{Alekseev:2004np} that the so-called
Dorfman bracket naturally appears in the Poisson-bracket of currents
of a two-dimensional sigma-model. The Dorfman bracket is an example
of a derived bracket and it is possible to obtain more general derived
brackets in a similar manner, although in a somewhat different setting
as in \cite{Alekseev:2004np}. The present contribution is intended
to give an idea where this relation comes from. For further references
and details we refer to \cite{Guttenberg:2006zi}. \enlargethispage*{1cm}

Consider first a one-dimensional sigma model (point particle) with
the same number of anticommuting variables $\ce^{m}$ as of commuting
target space coordinates $x^{m}$. This can be either a topological
theory (with $\ce^{m}$ being ghosts) or a theory with worldline supersymmetry.
Let $p_{m}$ and $\be_{m}$ be the corresponding conjugate momenta
in the Hamiltonian formalism. For quantization we choose the ordering
where all momenta are moved to the right and then replace $p_{m}\to\frac{\hbar}{i}\partl{x^{m}}$
and $\be_{m}\to\frac{\hbar}{i}\partl{\ce^{m}}$. Phase space functions
$F$ thus become operators $\hat{F}$ which will act on wave functions.
The latter are functions of $x^{m}$ and $\ce^{m}$. This setting
has a nice geometrical interpretation if one identifies $\ce^{m}$
and $\be_{m}$ with the coordinate basis elements of cotangent and
tangent space of the target space: $\ce^{m}\cong\de x^{m},\qquad\be_{m}\cong\pe_{m}$.
The wave functions are then functions of $x^{m}$ and $\de x^{m}$
or, in other words, formal sums of differential forms. Phase space
functions (before quantization) which do not depend on $p_{m}$ are
just formal sums of multivector valued forms. Vector fields in particular
correspond to phase space functions of the form $v(x,\be)=v^{m}(x)\be_{m}$.
Quantization embeds the vector fields into the space $\mbox{End}(\Omega^{\bullet}T^{*}M)$
of operators acting on forms. The result is nothing but the interior
product $\hat{v}=\frac{\hbar}{i}\ip_{v}$. 

For phase space functions $K^{(k,k')}(x,\ce,\be)$ which are of $k$-th
power in $\ce^{m}$ and of $k'$-th power in $\be_{m}$ (multivector
valued forms of form degree $k$ and multivector degree $k'$) one
arrives similarly at $\hat{K}=\left(\frac{\hbar}{i}\right)^{k'}\ip_{K}$.
Here $\ip_{K}$ is a natural generalization of the interior product,
where all multivector indices are contracted with form-indices of
the differential form one is acting upon, and all the remaining indices
are antisymmetrized such that the result is a form again. \enlargethispage*{1cm}
One can also view the above relation as definition of $\ip_{K}$.
The graded commutator induces an algebraic bracket $[K,L]^{\Delta}$
on the multivector valued forms via \begin{equation}
[\ip_{K},\ip_{L}]=\ip_{[K,L]^{\Delta}}\label{eq:algebraic-bracket}\end{equation}
This algebraic bracket can be expanded in the number $p\geq1$ of
contracted indices between $K$ and $L$: $[K,L]^{\Delta}=\sum_{p\geq1}[K,L]_{(p)}^{\Delta}$,
s.t. $[K,L]_{(p)}^{\Delta}$ has form degree $k+l-p$ and multivector
degree $k'+l'-p$. The term of highest form degree or multivector
degree (at $p=1$) is itself an algebraic Lie bracket (called big
bracket) as it is simply the Poisson bracket between $K$ and $L$:\\
$[K,L]_{(1)}^{\Delta}=\{ K,L\}$.

While $K(x,\ce,\be)$ corresponds after the embedding (or quantization)
to an algebraic operator, phase space functions that contain the variable
$p_{m}$ rather correspond to differential operators, because $p_{m}$
acts as $x$-derivative after quantization. For a phase space function
$F^{(f,f',f'')}(x,\ce,\be,p)$ which is a monomial of degree $f,f'$
and $f''$ in $\ce^{m}$, $\be_{m}$ and $p_{m}$ respectively, it
is now natural to define a generalization of the interior product
via $\ip_{F}\equiv\left(\frac{i}{\hbar}\right)^{f'+f''}\hat{F}$.
This implies a generalization of the algebraic bracket (\ref{eq:algebraic-bracket})
of multivector valued forms to arbitrary phase space functions. In
addition, the exterior derivative can now be written as an {}``interior
product'': $\de=\ip_{\bs{{\scriptstyle c}}^{m}p_{m}}$. Due to (\ref{eq:algebraic-bracket}),
its graded commutator with another interior product is again an interior
product which will induce a differential on phase space functions
\cite{Guttenberg:2006zi}. \begin{equation}
\left[\de,\ip_{K}\right]\equiv\ip_{\de K}\label{eq:extended-d}\end{equation}
This differential is just an extension of the exterior derivative
(originally only defined on forms, i.e. functions of $x^{m}$ and
$\ce^{m}$ only) to arbitrary phase space functions. It boils down
to fixing $\de(\be_{m})=p_{m},\quad\de(p_{m})=0$ in addition to $\begin{array}{c}
\de(x^{m})=\ce^{m}\end{array}$ and $\de(\ce^{m})=0$. 

Apart from the commutator, there is another natural bracket in operator
space, namely the derived bracket \cite{Kosmann-Schwarzbach:2003en}
of the commutator by the de Rham differential \begin{equation}
[a,_{\bs{d}}b]\equiv[[a,\de],b]\quad\forall a,b\in\mbox{End}(\Omega^{\bullet}T^{*}M)\label{eq:derived-bracket}\end{equation}
Due to $[\de,\de]=0$, this bracket obeys a graded Jacobi identity,
but it is in general not skew-symmetric and therefore s.th. which
is called a Loday bracket. Via the embedding with the interior product,
it induces the vector Lie bracket on vector fields: $[\ip_{v},_{\de}\ip_{w}]=\ip_{[v\bs{,}w]}$.
Because of (\ref{eq:extended-d}) and (\ref{eq:algebraic-bracket}),
this can be written as $[\de v,w]^{\Delta}=[v\bs{,}w]$. The derived
bracket (\ref{eq:derived-bracket}) in operator space induces in the
same way a derived bracket on arbitrary phase space monomials $F^{(f,f',f'')}(x,\ce,\be,p)$
of the form \begin{equation}
[F\bs{,}G]=-(-)^{f+f'}[\de F,G]^{\Delta}\end{equation}
 with $[F\bs{,}G]$ defined as $[\ip_{F},_{\de}\ip_{G}]=\ip_{[F\bs{,}G]}$.
This bracket can be seen as a generalization of the vector Lie bracket,
although the result of the bracket will in general not be a multivector
valued form if we restrict $F$ and $G$ to be of that type. In the
special case of multivectors (including the vector field case of above),
however, the result is again a multivector and it is indeed just the
Schouten bracket. For the bracket between vector valued forms, the
derived bracket coincides only up to a total derivative with the so-called
Fr\"ohlicher-Nijenhuis bracket. 

\enlargethispage*{1cm} One can similarly define a derived bracket
of the big bracket as $[F\bs{,}G]_{(1)}\equiv-(-)^{f+f'}[\de F,G]_{(1)}^{\Delta}$.
It will be the leading order term in a quantum calculation of the
full bracket. In the case of vector valued forms and in the case of
multivectors this first term already coincides with the full derived
bracket. The same is true for formal sums of vectors and forms.

Our special field content allows us to introduce canonical conjugate
{}``superfields'' $\begin{array}{c}
\Phi^{m}(\tet)\equiv x^{m}+\tet\ce^{m}\end{array}$ and $\Es_{m}(\tet)\equiv\be_{m}+\tet\be_{m}$, where $\tet$ is some
auxiliary Grassmann variable. The exterior derivative is now clearly
implemented by $\partial_{\tet}$. Let $K(\tet)\equiv K(\Phi,\partial_{\tet}\Phi,\Es)$
be a multivector valued form as before, but with the arguments replaced
by the corresponding {}``superfields'': $x^{m}\To\Phi^{m},\quad\ce^{m}\To\partial_{\tet}\Phi^{m},\quad\be_{m}\To\Es_{m}$.
The almost trivial relation between Poisson bracket and big bracket
of before then changes in an interesting manner:\begin{eqnarray}
\lqn{\{ K(\tet'),L(\tet)\}=}\nonumber \\
 & = & \delta(\tet'-\tet)[\de K,L]_{(1)}^{\Delta}(\tet)+[K,L]_{(1)}^{\Delta}(\tet)\label{eq:result}\end{eqnarray}
In other words,\enlargethispage*{1cm}  the derived bracket naturally
appears as coefficient of the delta function while the algebraic bracket
appears as coefficient of the derivative of the delta function ($\partial_{\tet}\delta(\tet-\tet')=1$).
Going to higher worldvolume dimensions with coordinates $\sigma^{\mu}$
does not spoil this result as long as no $\sigma$-derivatives are
involved. The righthand side is simply furnished by an additional
delta function in the worldvolume coordinates. At first sight it is
less obvious that a very similar relation holds if one completely
interchanges the role of $\sigma$ and $\tet$. This works particularly
well in the Lagrangian formalism using antifields instead of the conjugate
momenta and the antibracket instead of the Poisson bracket. The result
in (\ref{eq:result}) and its antibracket-equivalent were applied
in \mbox{\cite{Guttenberg:2006zi}} to a so-called generalized complex
structure, which is a formal sum of a 2-form, a 2-vector and of a
vector-valued form. The relations of certain 2-dimensional sigma-models
to integrability of generalized complex structures as they were discussed
among others in \cite{Zabzine:2005qf,Zucchini:2004ta} become particularly
transparent in this way.

\end{document}